\documentclass[a4paper,11pt]{article}
\pdfoutput=1

\usepackage{jcappub}
\usepackage{subcaption}

\title{Axion star condensation\\ around primordial black holes\\ and microlensing limits}

\author{Ziwen Yin}
\author{and Luca Visinelli}

\affiliation{Tsung-Dao Lee Institute (TDLI),
No.\ 1 Lisuo Road, 
 201210 Shanghai, China;}
\affiliation{School of Physics and Astronomy, Shanghai Jiao Tong University,
800 Dongchuan Road, 200240 Shanghai, China}

\emailAdd{ziwenyin@sjtu.edu.cn}
\emailAdd{luca.visinelli@sjtu.edu.cn}

\abstract{We present novel findings concerning the parameter space of axion stars, extended object forming in dense dark matter environments through gravitational condensation. We emphasize their formation within the dense minihalos that potentially surround primordial black holes and in axion miniclusters. Our study investigates the relation between the radius and mass of an axion star in these dense surroundings, revealing distinct morphological characteristics compared to isolated scenarios. We explore the implications of these results when applied to the bound state between a primordial black hole and an axion star and the gravitational microlensing from extended objects, leading to insights on the observational constraints from such ``halo'' axion stars. We provide a constraint on the fraction of the galactic population of axion stars from their contribution to the microlensing events from the EROS-2 survey, using the numerical resolution of the Schr\"odinger-Poisson equation.}


\begin{document}
\maketitle
\flushbottom

\section{Introduction}
\label{sec:introduction}

Astrophysical and cosmological observations suggest that approximately 27\% of the energy content of the universe resides in dark matter (DM)~\cite{Planck:2018vyg, Planck:2018lbu, ACT:2023kun, DES:2022urg}. This picture agrees with structure formation~\cite{eBOSS:2020yzd}, the gravitational lensing of distant objects~\cite{Natarajan:2017sbo, Meneghetti:2020yif}, and the rotation curves of galaxies~\cite{Rubin:1980zd}. The gravitational effects of DM are undeniable, including its importance in the cosmological history. Yet, its elusive non-gravitational interactions have remained beyond the reach of traditional observational methods. For this, the nature of DM is currently a debated subject, with various theories proposing a particle candidate that stems beyond the Standard Model of particles (SM).

One compelling DM candidate is the QCD axion~\citep{Weinberg:1977ma, Wilczek:1977pj}, a hypothetical particle predicted within the solution to the strong-CP puzzle proposed by Peccei and Quinn (PQ)~\cite{Peccei:1977hh}. The QCD axion's origin lies within the framework of the PQ mechanism, which introduces a new global symmetry to QCD, thereby dynamically suppressing CP-violating terms. The breaking of this symmetry at high energies gives rise to a pseudo Nambu-Goldstone boson, the axion, endowed with a mass that is inversely proportional to the energy scale at which the symmetry breaking occurs. For recent reviews see refs.~\cite{DiLuzio:2020wdo, Chadha-Day:2021szb}. Remarkably, the QCD axion might serve both as a solution to the strong-CP puzzle while concurrently offering a plausible explanation for the cosmic DM budget. In fact, the particle's lifetime would exceed the age of the Universe, behaving as a pressureless non-interacting fluid on galactic or cosmological scales~\cite{Preskill:1982cy, Abbott:1982af, Dine:1982ah}.

Along with the QCD axion, other similar axion-like particles (ALPs) have also been proposed in extensions of the SM~\cite{Georgi:1986df, Svrcek:2006yi, Conlon:2006tq, Arvanitaki:2009fg, Arias:2012az, Visinelli:2018utg, Gendler:2023kjt}. Although ALPs do not directly solve the strong-CP problem, they share similar properties with the QCD axion such as their nature as a pseudo-scalar Goldstone boson. Here, we limit ourselves to considering the QCD axion only, although several studies have been devoted to the morphologies and indirect effects of macroscopic structures made of ALPs.

The properties of the primordial distribution of the axion field are affected by its cosmological history, with large inhomogeneities developing in the scenario where the PQ symmetry breaks after inflation~\cite{Beltran:2006sq, Hertzberg:2008wr, Visinelli:2009zm, Visinelli:2009kt}. This has been recently confirmed in numerical simulations of the axion field evolution~\cite{Hiramatsu:2012gg, Klaer:2017ond, Gorghetto:2018myk, Buschmann:2019icd, Gorghetto:2020qws, Buschmann:2021sdq, Hoof:2021jft, Saikawa:2024bta}. Owing to this clumpy spatial distribution, it is expected for many regions to undergo gravitational collapse as fluctuations become non-linear, leading to axion miniclusters~\cite{Hogan:1988mp, Kolb:1993hw, Kolb:1993zz, Kolb:1994fi, Zurek:2006sy}. These objects are expected to form if the PQ symmetry breaks after inflation, as shown in both theoretical setups~\cite{Visinelli:2018wza, Ellis:2022grh} and numerical simulations~\cite{Vaquero:2018tib, Eggemeier:2019khm, Xiao:2021nkb, Ellis:2022grh}. These dense objects might even survive tidal disruption to date and be part of the galactic halos~\cite{Kavanagh:2020gcy, Shen:2022ltx, OHare:2023rtm, Maseizik:2024qly}, leading to indirect evidences such as microlensing~\cite{Fairbairn:2017sil, Croon:2020wpr, Croon:2020ouk, Schiappacasse:2021zlr} and particle conversion in strong magnetic fields~\cite{Edwards:2020afl, Witte:2022cjj}. 

These inhomogeneities could be expected even in the complementary scenario where the PQ symmetry breaks before or during inflation, through the formation of axionic ``dark minihalos'' around primordial black holes (PBHs)~\cite{Hertzberg:2020hsz}. The formation of these structures could proceed even in the absence of primordial DM seeds. The importance of this remark stems in the fact that PBHs could have formed at different mass scales and across various cosmological scenarios, possibly independently of the history of the axion field.

In addition to its role as a potential DM constituent, the QCD axion has also been implicated in the formation of axion stars, time-periodic solutions of the Einstein-Klein-Gordon equations~\cite{Seidel:1991zh, Seidel:1993zk}. Axion stars are hypothesized to arise from the gravitational collapse of axion clouds, wherein the gravitational collapse is counteracted by the pressure term arising from Heisenberg's uncertainty principle, thus leading to a stable, gravitationally-bound configuration. For this, axion stars exhibit intriguing characteristics, including a Bose-Einstein condensate-like behavior, wherein vast numbers of axions occupy the lowest energy state, manifesting macroscopic quantum phenomena on macroscopic scales. Axion stars might have formed inside the dense environment of axion miniclusters or the dense PBH minihalos~\cite{Levkov:2016rkk, Eggemeier:2019jsu, Hertzberg:2020hsz, Dmitriev:2023ipv}. The process of nucleation has been assessed numerically with the corresponding gravitational timescale being under control from the theory viewpoint.\footnote{Related configurations called axitons and corresponding to oscillating solutions of the Klein-Gordon equation have also been recovered in simulations of the axion field in the early Universe~\cite{Kolb:1994fi, Vaquero:2018tib}.} Axion stars have been extensively studied as isolated objects in the vacuum, in the so-called ``dilute'' regime~\cite{Seidel:1991zh, Chavanis:2011zm, Braaten:2015eeu, Visinelli:2017ooc, Schiappacasse:2017ham}.

Several techniques employed in the search for diffuse dark clumps in galaxies include parametric-induced explosions of axion stars into photons~\cite{Hertzberg:2018zte, Carenza:2019vzg, Amin:2020vja} or relativistic axions~\cite{Eby:2015hyx, Levkov:2016rkk, Escudero:2023vgv}, the gravitational wave~\cite{Chung-Jukko:2024hod} and photon resonances~\cite{Hertzberg:2018zte} from post-mergers, and gravitational microlensing. This latter method is characterized by the amplification of brightness in background source stars as clumps pass near the line of sight. Notably, the abundances of massive compact halos and PBHs undergo stringent scrutiny through gravitational lensing observations. Surveys such as the Exp\'erience de Recherche d’Objets Sombres (EROS)~\cite{Palanque-Delabrouille:1997cxg, EROS-2:2006ryy}, the MAssive Compact Halo Objects (MACHO)~\cite{MACHO:2000qbb}, the Optical Gravitational Lensing Experiment (OGLE)~\cite{Wyrzykowski:2011tr, Mroz:2024mse}, and the Microlensing Observations in Astrophysics (MOA)~\cite{2001MNRAS.327..868B} collaborations play pivotal roles in these constraint efforts. See ref.~\cite{Carr:2020gox} for a recent review on the current status of observational constraints on PBHs.

Unlike PBHs, DM clumps in the form of dark minihalos and axion stars possess internal structures that generally make them unsuitable for a treatment in terms of point-like massive objects, when dealing with microlensing events. Consequently, to accurately assess the microlensing constraints on axion clumps, finite lens size effects have to be accounted for. Numerous investigations have explored microlensing events caused by astrophysical objects with finite extents, including boson stars and self-similar subhalos~\cite{Croon:2020ouk, Croon:2020wpr, Oguri:2022fir, CrispimRomao:2024nbr, Croon:2024rmw}. Additionally, the gravitational lensing phenomenon associated with axion miniclusters has been explored in previous research~\cite{Kolb:1995bu, Fairbairn:2017sil, Fujikura:2021omw}.

In this work, we consider the formation of an axion star inside a dark minihalo surrounding a PBH and in axion miniclusters. The density profile of the minihalo is predicted from the spherical collapse and accretion of pressureless DM. The conditions imposed in this dense environment modify the morphology of the axion star, leading to a different relation between its mass and radius compared to the case of a dilute star. We call these objects ``halo'' axion stars. We consider light PBHs in the asteroid mass window, where axion stars of comparable magnitude in mass can form through gravitational condensation. These objects lead to interesting phenomenology such as microlensing from extended objects, thus providing a complementary route to search for both a PBH population and the DM component in the form of a light boson. One of the primary objectives of this study is to derive microlensing constraints from surveys targeting axion clumps. Since the model predicts an extended lens in the form of axion stars and dark minihalos, we constrain the fraction of these objects with microlensing using the data from the EROS-2 survey~\cite{EROS-2:2006ryy}. We use the full numerical solution of the Schr\"odinger-Poisson system for modeling the profile of the axion star.

The paper is organized as follows. In section~\ref{sec:methods} we introduce the methods used in our computations. Results are drawn in section~\ref{sec:results} and a discussion of gravitational microlensing is carried out in section~\ref{sec:discussion}. We comment on the outcome and future directions in the conclusions of section~\ref{sec:conclusions}. We set $\hbar = c = 1$ throughout the paper.

\section{Methods}
\label{sec:methods}

\subsection{Properties of the dark minihalo}

The process of modeling DM accretion around one PBH has been tackled both analytically and through N-body simulations. Here, PBHs constitute only a very small fraction of the DM, with the axion contributing to the bulk DM budget. Some of the axion DM accumulates around the PBH, forming a dark halo~\cite{Bringmann:2011ut, Berezinsky:2013fxa, Berezinsky:2014wya, Adamek:2019gns, Serpico:2020ehh, Carr:2020mqm}.\footnote{The steepness of the profile $\propto -9/4$ has been first expected in a spherically-symmetric collapsed region within the Einstein-de Sitter metric~\cite{Fillmore:1984wk, Bertschinger:1985pd}.} This occurs in the early Universe through the joint action of the gravitational attraction from rogue PBHs and the Hubble expansion, which leads to the acceleration of a DM shell of radius $r$ around an isolated PBH of mass $M_{\rm PBH}$ as
\begin{equation}
    \label{eq:shell}
    \frac{{\rm d}^2r}{{\rm d}t^2} = -\frac{r_g}{r^2} + \left(\dot H+H^2\right)r\,,
\end{equation}
where $r_g = GM_{\rm PBH}$ is the gravitational radius of the PBH, $H$ is the Hubble rate, and $\dot H$ its variation with respect to time $t$. By the time of matter-radiation equality (``eq''), the dark minihalo around a PBH of mass $M_{\rm PBH}$ possesses a spiky distribution with density profile
\begin{equation}
    \label{eq:spike}
    \rho_{\rm spike}(r) = \frac{\rho_{\rm eq}}{2}\,\left(\frac{r_{\rm ta}(t_{\rm eq})}{r}\right)^{9/4}\,,
\end{equation}
where $\rho_{\rm eq}$ is the density at redshift $z_{\rm eq} \approx 3400$ and the time $t_{\rm eq}$ is fixed through the expression
\begin{equation}
	t_{\rm eq} = \left(\frac{3}{32 \pi G \rho_{\rm eq}}\right)^{1/2}\,.
\end{equation}
The turn-around radius $r_{\rm ta}$ defines the region within which the gravitational influence of the PBH overcomes cosmic expansion and corresponds to setting $\dot r=0$ in eq.~\eqref{eq:shell}, or~\cite{Adamek:2019gns}
\begin{equation}
	\label{eq:turnaround}
	r_{\rm ta}(t) \approx (2r_g\,t^2)^{1/3}\,.
\end{equation}
In this setup, the expression in eq.~\eqref{eq:spike} holds for $r < r_{\rm ta}(t_{\rm eq})$ at time $t_{\rm eq}$. For example, a PBH of mass $M_{\rm PBH} = 10^{-13}\,M_\odot$ has $r_{\rm ta}(t_{\rm eq}) \approx 0.26$\,au.

The dark minihalo radius keeps growing through accretion, as expected both from simulations~\cite{Mack:2006gz, Ricotti:2007au} as well as calculations of the virial mass and radius~\cite{Berezinsky:2013fxa}. The halo radius can be fixed by setting the cutoff at the halo overdensity $\delta_H = 1$,
\begin{equation}
    \label{eq:Rhalo}
    R_{\rm halo}(z) \approx \, r_{\rm ta}(t_{\rm eq})\, \left(\frac{1 + z_{\rm eq}}{1 + z}\right)^{4/3}\,.
\end{equation}
At any redshift, the mass of the dark minihalo can be express by integrating the density profile to the radius $R_{\rm halo}(z)$,
\begin{equation}
    M_{\rm halo}=\int_0^{R_{\rm halo}}{\rm d}r\,4\pi r^2\rho_{\rm spike}(r)= \frac{8\pi}{3}\rho_{\rm eq}r^3_{\rm ta}(t_{\rm eq})\left(\frac{R_{\rm halo}(z)}{r_{\rm ta}(t_{\rm eq})}\right)^{3/4} \approx \frac{8}{9}M_{\rm PBH}\,\left(\frac{1+z_{\rm eq}}{1+z}\right)\,,
\end{equation}
where the additional factor 8/9 accounts for the evolution in the matter-dominated era. This result, previously obtained with other methods~\cite{Mack:2006gz, Ricotti:2007au}, holds down to the redshift $z_{\rm gal} \sim 30$ where growth typically stops due to tidal interactions. At any time, the mass enclosed within the radius $r$ and the corresponding virial velocity $v$ can be obtained from integrating to the radius $r < R_{\rm halo} \equiv R_{\rm halo}(z_{\rm gal})$ within the DM halo,
\begin{eqnarray}
    M_{\rm encl}(r) &=& M_{\rm PBH} + M_{\rm halo}\, \left(\frac{r}{R_{\rm halo}}\right)^{3/4}\,,\label{eq:enclosedmass}\\
    v^2 &=& \frac{GM_{\rm encl}(r)}{r}\,.
\end{eqnarray}

\subsection{Axion star nucleation}

The results of the previous section are obtained for a pressureless fluid and thus apply to different models of cold DM. We now specialize the discussion to the QCD axion, where particle-particle interactions can lead to an enhancement of the quantum properties of these light bosons through gravitational condensation.

The condensation of axion stars inside miniclusters and dark minihalos has been studied in the literature through numerical simulations, accounting for both self-gravitating light particles~\cite{Levkov:2018kau} and with the inclusion of self-interactions~\cite{Kirkpatrick:2020fwd, Kirkpatrick:2021wwz, Chen:2021oot, Chen:2023bqy, Dmitriev:2023ipv, Jain:2023tsr}. If gravity provides the dominant contribution to the particle interaction, the time scale of the Bose star formation process is~\cite{Levkov:2018kau}
\begin{equation}
    \label{eq:gr}
    \tau_{\rm gr} \approx \frac{\sqrt{2}}{12\pi^3}\frac{m_a v^6}{G^2n^2\Lambda}\,,
\end{equation}
where $n$ is the DM number density and the Coulomb logarithm $\Lambda = \ln (m_a vR_{\rm halo})$ accounts for the scale of the system $R_{\rm halo}$. While the quantities $v$ and $n$ entering eq.~\eqref{eq:gr} involve local properties of the halo, the size of the halo only appears in the logarithmic correction, so that the structure of the halo does not greatly affect the outcome of the computation.

We demand two conditions that should be satisfied inside the dark minihalo for the formation of an axion star to take place. First of all, the halo has to be sufficiently large to host the de Broglie wavelength of the axion,
\begin{equation}
    \label{eq:debroglie}
    m_a \, v \, R_{\rm halo} \gg 1\,,
\end{equation}
and the kinetic regime condition should occur sufficiently fast~\cite{Levkov:2018kau}
\begin{equation}
    \label{eq:kinetic}
     m_a\,v^2\,\tau_{\rm gr} \gg 1\,.
\end{equation}
We also demand that the axion star actually forms on a timescale shorter than the age of the Universe at redshift $z$,
\begin{equation}
    \label{eq:formationtime}
    \Gamma_{\rm kin}(r)\gtrsim \left(\frac{t_0}{(1+z)^{3/2}}\right)^{-1}\,,
\end{equation}
where the kinetic relaxation rate is~\cite{Hertzberg:2020hsz}
\begin{equation}
    \Gamma_{\rm kin}(z) \approx n\,\sigma_{\rm gr}\,v\,N \approx n^2\,\frac{48\pi^2 G^2\Lambda}{m_a\,v^6}\,.    
\end{equation}
Setting $n=\rho_{\rm spike}(r)/m_a$, eq.~\eqref{eq:formationtime} at a given redshift $z$ can be used to obtain the critical radius $R_{\rm crit}$ within which at least one axion star forms inside the dark minihalo,
\begin{equation}
    \label{eq:Rcrit}
    R_{\rm crit}=r_{\rm ta}\left(\frac{81\Lambda}{128\pi G\rho_{\rm eq}r^6_{\rm ta}(t_{\rm eq})m_a^3}\frac{t_0}{(1+z)^{3/2}}\right)^{4/15}\,,
\end{equation}
with the additional requirement $R_{\rm crit} < R_{\rm halo}(z)$.

Under the condition for the nucleation outlined above, we proceed to understand the properties of nonrelativistic axion stars, solitonic solutions of the Schr{\"o}dinger-Poisson equation~\cite{Seidel:1991zh}
\begin{align}
    \nabla^2\phi&=4\pi G\left(\rho+M_{\rm PBH}\delta^{(3)}({\bf r})\right)\,,\label{eq:poisson} \\
    \epsilon\psi&=-\frac{1}{2m}\nabla^2\psi+\left(\frac{\lambda_\Phi}{8m^2}\psi^*\psi+ m\phi\right)\psi\,.\label{eq:quantum}
\end{align}
Here, $\lambda_\Phi$ is the self-coupling constant, see eq.~\eqref{eq:schroedinger} in Appendix~\ref{sec:structureAS}, $\phi$ is the gravitational potential of the axion star and the PBH source, and the relation between energy density $\rho$ and $\psi$ is
\begin{equation}
    m|\psi|^2=\rho\,.
\end{equation}
Axion stars are expected to form in the central regions inside the dark minihalo surrounding a PBH and in axion miniclusters. If the nucleation occurs far from the center, the gravity potential of the PBH does not enter the Poisson eq.~\eqref{eq:poisson} but it would affect the structure of the axion star through tidal interactions~\cite{Hertzberg:2020hsz}. The mechanism proceeds if the central density of the minihalo is sufficiently high for two-to-two processes to relax the energy of the axions in the inner core, leading to particle condensation~\cite{Levkov:2018kau, Kirkpatrick:2020fwd, Kirkpatrick:2021wwz, Chen:2021oot, Chen:2023bqy, Dmitriev:2023ipv, Jain:2023tsr}. For a halo of ultralight particles, numerical simulations predict the relation between the halo mass and the axion star mass as~\cite{Schive:2014dra, Schive:2014hza}
\begin{equation}
    \label{eq:axionstarmass}
    M_{\rm AS} \approx 1.8\times 10^{-10}\,M_\odot\,\left(\frac{1+z}{1+z_{\rm eq}}\right)^{1/2}\,\left(\frac{\mu \rm eV}{m_a}\right)\,\left(\frac{M_{\rm halo}}{M_\odot}\right)^{1/3}\,.
\end{equation}
Several independent simulations with N-bodies on the formation of a solitonic core in massive halos find similar results for the cored behaviour, although there is still disagreement on the specific relation between the core mass and the halo mass~\cite{Schwabe:2016rze, Mocz:2017wlg, Nori:2020jzx}. Moreover, the soliton core might be subject to tidal stripping that alters its shape~\cite{Chan:2021bja}. The profile of a solitonic core around a supermassive BH has also been studied numerically~\cite{Bar:2018acw, Davies:2019wgi}. Consistency demands that the mass of the axion star be $M_{\rm AS} < M_{\rm halo}$, leading to a fourth requirement that has to be satisfied for the formation of the star in the halo.

\section{Results}
\label{sec:results}

\subsection{Dark minihalo}

We now specialize these requirements to the dark minihalo discussed previously. Setting $\Lambda \approx 7$, combining the kinetic condition in eqs.~\eqref{eq:debroglie},~\eqref{eq:kinetic},~\eqref{eq:formationtime} and~\eqref{eq:axionstarmass} gives
\begin{eqnarray}
    m_a v R_{\rm halo} &\sim& 500\left(\frac{m_a}{10^{-5}{\rm\,eV}}\right)\left(\frac{10^3}{1+z}\right)^{\frac{7}{6}}\left(\frac{M_{\rm PBH}}{10^{-13}M_{\odot}}\right)^{\frac{2}{3}}\,,\label{eq:halosize}\\
    (m_a v^2 \tau_{\rm gr})^{1/4} &\sim& 400\left(\frac{m_a}{10^{-5}{\rm\,eV}}\right)\left(\frac{10^3}{1+z}\right)^{\frac{7}{6}}\left(\frac{M_{\rm PBH}}{10^{-13}M_{\odot}}\right)^{\frac{2}{3}}\,,\\
    \left(\frac{R_{\rm halo}}{R_{\rm crit}}\right)^{\frac{5}{4}} &\sim& 500\,\left(\frac{m_a}{10^{-5}{\rm\,eV}}\right)\left(\frac{10^3}{1+z}\right)^{\frac{7}{6}}\left(\frac{M_{\rm PBH}}{10^{-13}M_{\odot}}\right)^{\frac{2}{3}}\,,\label{eq:boundRcrit}\\
    \frac{M_{\rm halo}}{M_{\rm AS}}&\sim&500\,\left(\frac{m_a}{10^{-5}{\rm\,eV}}\right)\left(\frac{10^3}{1+z}\right)^{\frac{7}{6}}\left(\frac{M_{\rm PBH}}{10^{-13}M_{\odot}}\right)^{\frac{2}{3}}\,.\label{eq:ASbound}
\end{eqnarray}
Surprisingly, all four of the expressions above lead to the same functional dependence in terms of redshift $z$, axion mass $m_a$, and halo mass. This has not been anticipated before, since eq.~\eqref{eq:axionstarmass} has been derived from the numerical properties of the cosmological simulation and is thus independent from the considerations that lead to the first three expressions.

\begin{figure}[htbp]
    \centering
    \includegraphics[width=0.7\linewidth]{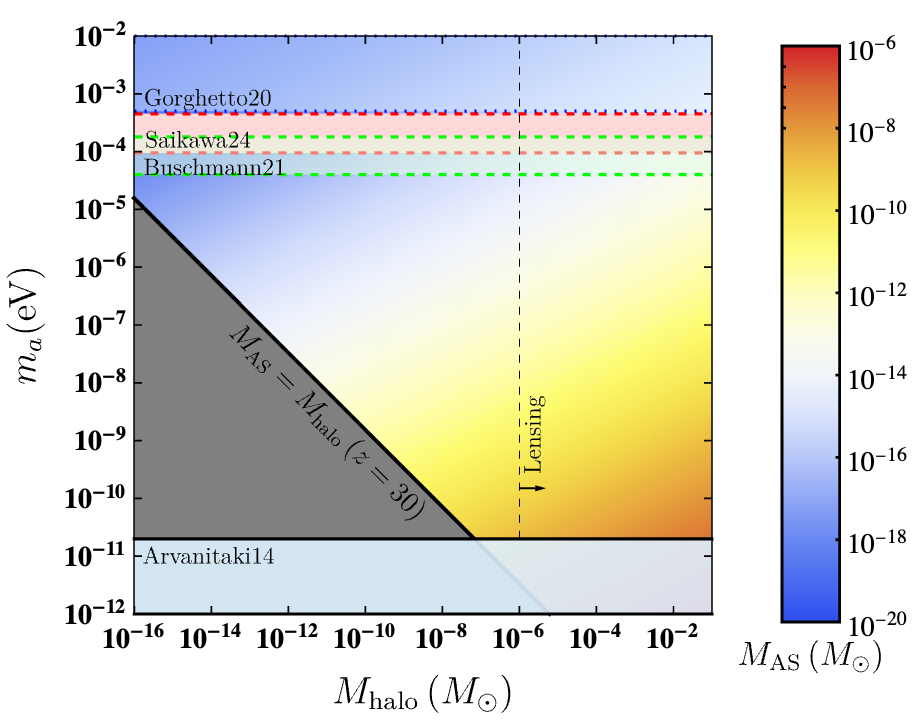}
    \caption{The mass of the axion star as a function of the halo mass (horizontal axis) and the QCD axion mass (vertical axis). The color map shows the mass of the corresponding axion star for values of the pair $\left(M_{\rm halo},m_a\right)$ at $z=30$. The region below the black solid line shows the constrain from the mass ratio between axion stars and the axion halo given in eq.~\eqref{eq:ASbound}. Also shown is the reach from gravitational microlensing (right of the vertical dashed line), the region excluded by stellar BH spin measurements (light blue), and the prediction of the QCD axion mass from numerical simulations (see text for details).}
    \label{fig:MASdensityplot}
\end{figure}

We impose that all four quantities above exceed unity, to satisfy the conditions for the formation and the stability of the axion star. When plotting these constraints on the parameter space $\left(M_{\rm halo}, m_a\right)$, the four expressions lead to parallel lines, with the latter coming from the requirement $M_{\rm AS} < M_{\rm halo}$ yielding the slightly most stringent bound. Figure~\ref{fig:MASdensityplot} summarizes the results obtained. The density plot accounts for the axion star mass in eq.~\eqref{eq:axionstarmass} as a function of the halo mass (horizontal axis) and the axion mass (vertical axis). Also shown is the bound derived from demanding that the axion star forms by $z\approx 30$, with a halo that is massive enough to contain it as $M_{\rm AS} < M_{\rm halo}$, as expressed in eq.~\eqref{eq:ASbound}. Also shown in the plot are the regions favored by some recent simulations of the QCD axion mass from cosmic string simulations, including ``Saikawa24''~\cite{Saikawa:2024bta} predicting $m_a\in [95\textrm{-}450]\,\mu$eV (area within the red dashed lines), ``Buschmann21''~\cite{Buschmann:2021sdq} predicting $m_a\in [40\textrm{-}180]\,\mu$eV (area within the green dot-dashed lines), and ``Gorghetto20''~\cite{Gorghetto:2020qws} predicting $m_a > 500\,\mu$eV (area above the blue dotted line). The region labeled ``Arvanitaki15'' marks the exclusion of the QCD axion in the mass range $m_a\in [6\textrm{-}200]\times 10^{-13}\,$eV, obtained from stellar black hole (BH) spin measurements~\cite{Arvanitaki:2014wva}. Finally, the region to the right of the vertical dashed line labeled ``Lensing'' can be probed by gravitational microlensing of the dark minihalos in the Galaxy, see section~\ref{sec:discussion} below. This result extends the findings in ref.~\cite{Hertzberg:2020hsz} to include the bounds presented in eqs.~\eqref{eq:halosize}-\eqref{eq:ASbound} in a systematic way, as well as providing a link between the parameter space and the mass of the axion star through eq.~\eqref{eq:axionstarmass}.

We consider now axion stars orbiting a PBH of mass $M_{\rm PBH} \gg M_{\rm AS}$ and nucleating close to the center of the distribution. The orbiting suggests that either the PBH or the axion star formed away from the center of the halo, so that a separation between these two objects would lead to the formation of a bound structure. Both these situations are in principle possible to obtain within the minihalo environment. The simulation of the axion condensation in ref.~\cite{Eggemeier:2019jsu} reports one of the halo consisting of two local density maxima, each producing one axion star. As this has occurred in a relatively small sample of halos, extended search on this frontier could result in even more exotic configurations. We briefly discuss the evolution of off-centered PBHs at the end of the section.

We generally expect that these axion stars possess an inner structure which is dominated by the gravitational potential of the PBH itself. In fact, when the self-gravity potential of the axion star density can be neglected, eq.~\eqref{eq:quantum} resembles a gravitational atom, with ground state
\begin{equation}
    \psi(r) = \alpha^3\,\left(\frac{M_{\rm AS}}{\pi\,r_g^3}\right)^{1/2}\,\exp\left(-\alpha^2 r/r_g\right)\,,
\end{equation}
where $\alpha = r_gm_a \equiv GM_{\rm PBH}m_a$ measures the gravitational coupling of the axion in the gravitational field of the PBH. The corresponding energy of the ground state is $E_0 = -\alpha^2m/2$, and the radius containing 90\% of the mass is
\begin{equation}
    \label{eq:RAS_PBH}
    R_{\rm AS} \approx 2.66\,r_g/\alpha^2\,.
\end{equation}
For the axion star radius to be relevant in this context, we generally expect that it encompasses the corresponding Schwarzschild radius of the PBH, $R_{\rm AS} \gtrsim 2r_g$, or $\alpha \lesssim 1$. Setting $m_a \gtrsim 10^{-9}\,$eV for the QCD axion leads to $M_{\rm PBH} \lesssim 0.1\,M_\odot$.

We briefly comment on the possibility that the halo of mass $M_{\rm halo}$ accretes the inner PBH. The axions orbit the halo by spanning a range of eccentricities, so that radial orbits cross an inner PBH within the timescale
\begin{equation}
    \tau_{\rm diff} = 2\pi\,\left(G\,\rho_{\rm halo}\right)^{\!-1/2} \approx 10^8{\rm\,yr}\,\left(\frac{M_{\rm halo}}{10^{-10}\,M_\odot}\right)^{-1/2}\,\left(\frac{R_{\rm halo}}{10^{-3}{\rm\,pc}}\right)^{3/2}\,.
\end{equation}
The axions in orbits with eccentricities $1-\varepsilon \approx 0$ are quickly absorbed into the PBH within the timescale $\tau_{\rm diff}$. Assuming that the orbital eccentricities are uniformly distributed in the phase space, these axions contribute to a small fraction $\sim (r_g/R_{\rm halo})^2$ of the total halo mass, while the axions with non-zero angular momentum continue to populate in the halo. Axions moving on such high-eccentricity orbits would keep accreting the PBH at a much smaller rate, which can be estimated as~\cite{Bondi:1952ni}
\begin{equation}
    \label{eq:Bondi}
    \frac{{\rm d}M_{\rm PBH}/{\rm d}t}{M_{\rm PBH}} = \frac{4\pi\,G^2\,M_{\rm PBH}\, \rho_{\rm halo}}{v^3}\,,
\end{equation}
where $v$ is the axion velocity dispersion. This leads to a much longer timescale
\begin{equation}
    \label{eq:taudyn}
    \tau_{\rm dyn} \approx 4\times10^{11}{\rm\,yr}\,\left(\frac{M_{\rm halo}}{10^{-10}\,M_\odot}\right)^{1/2}\,\left(\frac{R_{\rm halo}}{10^{-3}{\rm\,pc}}\right)^{3/2}\,\left(\frac{10^{-15}\,M_\odot}{M_{\rm PBH}}\right)\,.
\end{equation}
While heavier PBHs could end up consuming lighter halos, those that populate the ``asteroid'' mass window could be surrounded by a population of axions to this date. The accretion within the axion star follows similarly as in Eq.~\eqref{eq:Bondi}, once an additional factor $w/v$ is included to account for the extremely small velocity dispersion of the condensate $w \sim 1/m_a R_{\rm AS}$. While the enhanced density would speed up the accretion rate, including such a small velocity dispersion increases the lifetime of the star due to this process.

The setup introduced allows us to discuss the off-center formation and subsequent sink of the PBH into the minihalo. When the PBH formation does not occur at the center of the halo, it would spiral inwards via dynamical friction. This evolution is expected for supermassive BHs~\cite{Ullio:2001fb, Volonteri:2005pn} and for stellar-mass BHs in a dark minihalo~\cite{Eda:2013gg}. In the case of interest here, the dynamical friction is exerted by the axions in the halo, leading to the timescale in Eq.~\eqref{eq:taudyn} for the sinking process. During this period, any axion star that has formed in the center would form a bound state with the light PBH, possible leading to a source of gravitational waves at relatively high frequencies that could appear in resonant cavities~\cite{Berlin:2021txa, Gatti:2024mde}, phononic metamaterials~\cite{Kahn:2023mrj}, or enhanced magnetic conversion~\cite{Ringwald:2020ist}.

In this scenario, the PBH dynamics could lead to the stripping of the axion star within a few free-fall timescales. Stripping occurs when the tidal interaction exceeds the self-gravity of the axion star, or
\begin{equation}
    \rho_{\rm AS}\,d^3 \lesssim M_{\rm PBH}\,,
\end{equation}
where $d$ is the distance between the two compact objects and $\rho_{\rm AS} \approx M_{\rm AS}/R_{\rm AS}^3$. The effect of tidal stripping is then important when the PBH reaches the Roche limit
\begin{equation}
    d_{\rm crit} \sim \left(\frac{M_{\rm PBH}}{\rho_{\rm AS}}\right)^{1/3} \approx 2\times10^7{\rm\,m}\,\left(\frac{M_{\rm PBH}}{10^{-10}\,M_\odot}\right)^{1/3}\,\,\left(\frac{M_{\rm AS}}{10^{-14}\,M_\odot}\right)^{-4/3}\,\,\left(\frac{m_a}{100{\rm\,\mu eV}}\right)^2\,.
\end{equation}

\subsection{Axion miniclusters}

We compare these results with the shape and mass window for the axion star formed inside an axion minicluster (``amc''), for which the density undergoing spherical collapse reaches the value~\cite{Kolb:1994fi}
\begin{equation}
    \label{eq:amcdensity}
    \rho_{\rm amc}(\delta) \approx 140\,\delta^3\,(1+\delta)\,\rho_{\rm eq}\,,
\end{equation}
where the overdensity parameter $\delta$ models the excess density in the minicluster with respect to the background. The radius of the minicluster $R_{\rm amc}$ is then fixed in terms of $\delta$ and the halo mass $M_{\rm halo}$, so that eq.~\eqref{eq:amcdensity} gives the average density of the minicluster. With this choice, the inner density profile is found as~\cite{Fairbairn:2017sil, OHare:2017yze}
\begin{equation}
    \rho(r) = \frac{1}{4}\,\rho_{\rm amc}(\delta)\,\left(\frac{R_{\rm amc}}{r}\right)^{9/4}\,,
\end{equation}
for $r < R_{\rm amc}$, and zero otherwise. For this, we wish to test the relation between the axion star mass $M_{\rm AS}$ and the halo mass $M_{\rm halo}$ given in eq.~\eqref{eq:axionstarmass} by solving the Schr\"odinger-Poisson system of eqs.~\eqref{eq:poisson}-\eqref{eq:quantum} to obtain the mass of the star. More in detail, we proceed with the resolution using the formulas outlined in Appendix~\ref{sec:structureAS} which, following closely ref.~\cite{Chavanis:2011zm}, consists in rewriting the set of eqs.~\eqref{eq:poisson}-\eqref{eq:quantum} into a dimensionless form that does not depend on the QCD axion mass or the energy scale. Once the solution for the density profile is obtained, the radius of the axion star is then fixed through a prescription that involves the total mass of the star~\cite{Kaup:1968zz, Ruffini:1969qy}. Contrary to the fermion star, the boson star does not have a defined radius which needs to be specified by some additional condition. Previous work considered the radius containing a percentage of the mass of the star~\cite{Chavanis:2011zm, Visinelli:2017ooc}. Here, we define the radius of the dilute axion star as the radius enclosing 90\% of its total mass, or
\begin{equation}
    \label{eq:ASradius99}
    R_{\rm AS} \approx \frac{6.84}{G\,m_a^2\,M_{\rm AS}}\,,
\end{equation}
where the details to fix the numerical value in the numerator are discussed in Appendix~\ref{sec:structureAS}. For a given $m_a$, we obtain a family of solutions in terms of a set of masses and radii ($M_{\rm AS}, R_{\rm AS}$), depending on the central density of the axion star. The density of the star decreases with decreasing mass, so that a lower bound on the axion star mass comes from demanding that the star fits inside the minihalo, $R_{\rm AS} < r_{\rm ta}(t_{\rm eq})$. This requirement assures that the density of the star at the stellar radius does not fall below the average DM density at the time of formation. On the other end, an upper bound exists because of the existence of self-interactions that destabilize the stellar equilibrium, and leads to the request (see e.g.\ ref.~\cite{Visinelli:2017ooc})
\begin{equation}
    M_{\rm AS} \lesssim 3\times 10^{-20}\,M_{\odot}\,({\rm eV}/m_a)^2\,,
\end{equation}
where the expression above implicitly assumes the ratio of the up and down quark masses $m_u/m_d = 0.48$. See ref.~\cite{Fox:2023xgx} for the implications for axion-like particles.

\begin{figure}[htbp]
    \centering
    \includegraphics[width=0.7\linewidth]{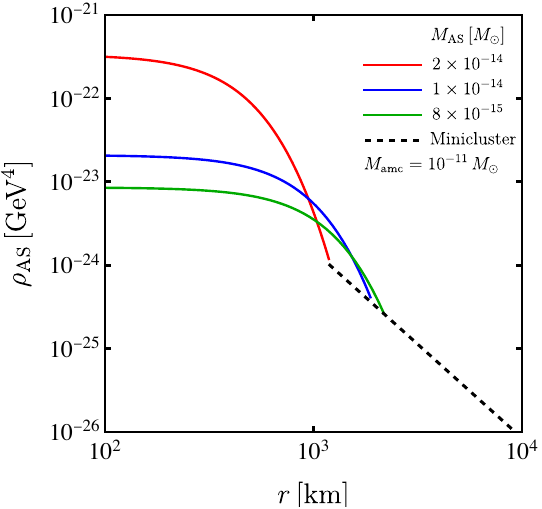}
    \caption{The density profile of various axion stars with different mass, hosted inside a minicluster of mass $M_{\rm amc} = 10^{-11}\,M_{\odot}$ with overdensity parameter $\delta = 1$. Specifically, we have fixed the masses $M_{\rm AS} = 2\times10^{-14}\,M_{\odot}$ (red), $M_{\rm AS} = 1\times10^{-14}\,M_{\odot}$ (blue), and $M_{\rm AS} = 8\times10^{-15}\,M_{\odot}$ (green). The mass of the QCD axion is fixed as $m_a=100\,\mu$eV.}
    \label{fig:profileplot}
\end{figure}

We compare the results obtained for a pristine ``dilute'' axion star formed in the vacuum with a star formed in the dense environment of a minicluster considered here. For such a ``halo'' axion star, we impose a different cutoff for its radius by considering the density of the surrounding environment. Assuming that the star forms at the center of the minicluster, the radius is obtained from the condition
\begin{equation}
    \label{eq:ASradius}
    \rho(R_{\rm AS}) = \frac{1}{4}\,\rho_{\rm amc}(\delta)\,\left(\frac{R_{\rm amc}}{R_{\rm AS}}\right)^{9/4}\,.
\end{equation}
The profiles for a dilute and a halo axion star are compared in figure~\ref{fig:profileplot} for the QCD axion mass $m_a = 100\,\mu$eV and an axion minicluster of mass $M_{\rm amc} = 10^{-11}\,M_{\odot}$ with overdensity parameter $\delta = 1$, for different values of the axion star. This corresponds to the radius $R_{\rm amc}\approx 5.4\times 10^7\,$km. The condensation at the core of the minicluster modifies the density profile and leads to a flat plateau instead of the spike predicted by the pressureless DM. This corresponds to the general behaviour seen in the numerical simulations that deal with structure formations with ultralight axions, although the length scales in play here are astronomical instead of galactic.

\begin{figure}[htbp]
    \centering
    \includegraphics[width=0.7\linewidth]{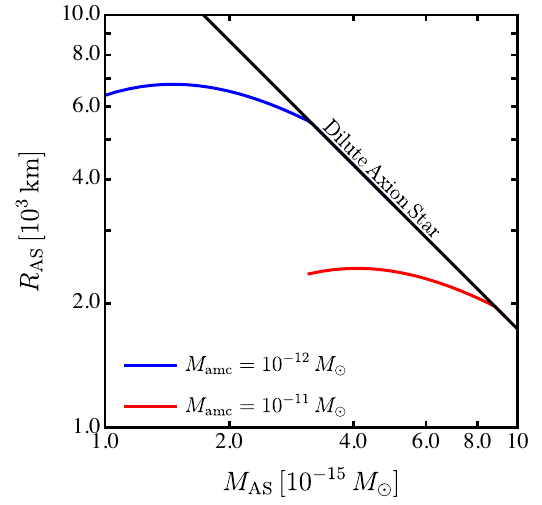}
    \caption{Mass-radius relation for different models of non-relativistic axion stars. We fix $m_a = 100\,\mu$eV. The results show the ``dilute'' branch (black dashed line), the ``halo'' axion star branch with $M_{\rm amc} = 10^{-12}\,M_\odot$ (blue line), and $M_{\rm amc} = 10^{-11}\,M_\odot$ (red line), both of overdensity parameter $\delta = 1$.}
    \label{fig:MRrelation}
\end{figure}

We now turn to fixing the radius of the axion star by means of the condition in eq.~\eqref{eq:ASradius}. The prescription leads to a non-trivial behaviour for mass-radius relation with respect to what is found in eq.~\eqref{eq:ASradius99}. This is shown in figure~\ref{fig:MRrelation} for the axion stars forming inside an axion minicluster of mass $M_{\rm amc} = 10^{-12}\,M_{\odot}$ (blue) or $M_{\rm amc} = 10^{-11}\,M_{\odot}$ (red), compared with the predictions from eq.~\eqref{eq:ASradius99} with $m_a = 100\,\mu$eV. A maximum radius is expected for the axion star, above which the condition for the condensation timescale in eq.~\eqref{eq:gr} would not be satisfied within the volume of the star. The prediction shown can be tested in simulations that follow the condensation of the axion star system as in refs.~\cite{Levkov:2018kau, Dmitriev:2023ipv}, however to date no study at various axion star masses exists.

One consequence of this finding is the impact on the mass distribution of axion stars. Assessing the halo mass function for axion miniclusters is a primary goal to correctly predict the outcome from indirect phenomena such as the gravitational microlensing from dense objects~\cite{Fairbairn:2017sil}. The halo mass function has been derived numerically in previous studies using hierarchical merging, revealing that such a function follows a steep power-law distribution~\cite{Enander:2017ogx, Vaquero:2018tib}. However, for relatively low masses of the miniclusters the gravitational condensation into an axion star has to be accounted for since, for such low-mass clusters, the timescale in eq.~\eqref{eq:gr} would lead to the formation of a large soliton. See e.g.\ ref.~\cite{Kavanagh:2020gcy} for a discussion of this effect in terms of the tidal stripping of miniclusters from nearby stars. The results in figure~\ref{fig:MRrelation} show that, on top of these considerations, the deviation of the axion star from the dilute regime and the appearance of a configuration with a maximum radius would also contribute to modifying the results obtained thus far, further motivating for a numerical investigation of this physical system in the coming future.

\section{Gravitational microlensing}
\label{sec:discussion}

Axion stars can be searched by various indirect methods such as gravitational microlensing. The search for DM in the form of faint compact objects has been theorized in refs.~\cite{1986ApJ...304....1P} and sparked the active search through gravitational microlensing by several surveys~\cite{Palanque-Delabrouille:1997cxg, EROS-2:2006ryy, MACHO:2000qbb, Wyrzykowski:2011tr, Mroz:2024mse, 2001MNRAS.327..868B}. For this, the lensing events from axion and boson stars has been extensively considered in previous literature, see e.g.\ refs.~\cite{Dabrowski:1998ac, Fairbairn:2017sil, Croon:2020ouk, Croon:2020wpr, Sugiyama:2021xqg, Fujikura:2021omw, Oguri:2022fir, CrispimRomao:2024nbr, Croon:2024rmw}.

We consider the lensing constraints on a population of dilute axion stars. We modify the results from previous work by considering the actual density distribution of the axion star inside the minihalo for a finite lens. This is achieved here by a numerical resolution of the density profile of the star using the Schr\"odinger-Poisson equation. We follow the standard notation in defining the observer-lens, lens-source, and observer-source distances as $D_{\rm L}$, $D_{\rm S}$, and $D_{\rm LS} \equiv D_{\rm S}-D_{\rm L}$, respectively, with the dimensionless ratio $x \equiv D_{\rm L}/D_{\rm S}$. Setting the angular radius of the Einstein ring $\theta_{\rm E}$, see eq.~\eqref{eq:rE}, gives the Einstein radius $R_{\rm E} \equiv D_{\rm L} \theta_{\rm E}$.

Using the axion stars formed in the dark minihalo as gravitational lenses with the profile obtained through eqs.~\eqref{eq:poisson}-\eqref{eq:quantum}, we derive the threshold impact parameter $u_{1.34}$ that leads to a total magnification $\mu_{\rm tot} \approx 1.34$, with the technical details given in Appendix~\ref{sec:microlensing}. Once the threshold impact parameter is provided, the total number of events is given in terms of the differential event rate as
\begin{equation}
    \label{eq:lensingevents}
    N_{\rm events} = E\,\int_0^1 {\rm d}x\int_{t_{\rm min}}^{t_{\rm max}}{\rm d}t_{\rm E}\,\frac{{\rm d}^2\Gamma}{{\rm d}x\,{\rm d}t_{\rm E}}\,,
\end{equation}
where $E$ is the total exposure time of the survey. The event rate per microlensing source of mass $M_{\rm lens}$ is given by~\cite{1991ApJ...366..412G} (see also eq.~(6) in ref.~\cite{Croon:2020ouk})
\begin{equation}
    \label{eq:lensingrate}
    \frac{{\rm d}^2\Gamma}{{\rm d}x\,{\rm d}t_{\rm E}} = \epsilon(2t_{\rm E})\frac{2D_{\rm S}}{v_0^2M_{\rm lens}}\rho_{\rm lens}(x)v_{\rm E}^4(x)\,\exp\left({-v_{\rm E}^2(x)/v_0^2}\right)\,,
\end{equation}
where $v_0 \approx 220{\rm\,km\,s^{-1}}$ is the galactic circular velocity. Here, $t_{\rm E}$ is the time required to cross the Einstein ring radius and $v_{\rm E}(x)= u_{1.34}R_{\rm E}(x)/t_{\rm E}$ is the velocity of the crossing. Finally, $\epsilon(2t_{\rm E})$ is the efficiency of the instrument.

\begin{figure}[htbp]
    \centering
    \includegraphics[width=0.7\linewidth]{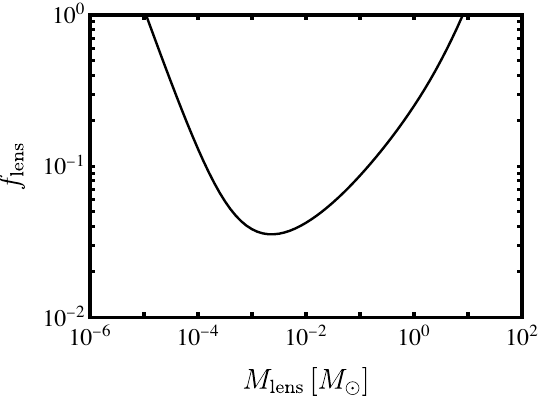}
    \caption{Lensing constraints on the mass of the lens using the data from the EROS-2 survey for a population of axion stars.}
    \label{fig:lensingbound}
\end{figure}

To obtain the bounds for the model we use the data from the EROS-2 survey~\cite{EROS-2:2006ryy} that includes observations of the Large Magellanic Cloud (LMC) and the Small Magellanic Cloud (SMC), to constrain the galactic fraction in axion stars. For this, the lensing event rate in eq.~\eqref{eq:lensingrate} is given in terms of the fraction of lensing objects $f_{\rm lens}$ as $\rho_{\rm lens}(r) = f_{\rm lens}\rho_{\rm DM}(r)$, where the DM distribution in the Milky Way halo is modeled with an isothermal profile as
\begin{equation}
    \rho_{\rm DM}(r) = \frac{\rho_s}{1+(r/r_s)^2}\,.
\end{equation}
Here, the normalization constant is taken as $\rho_s = 1.39{\rm\,GeV\,cm^{-3}}$ with the core radius $r_s = 4.38\,$kpc. The radius $r$ is related to the fractional distance $x$ between the Earth and the lens as
\begin{equation}
    r = \sqrt{r_\odot^2 + x^2D_S^2 - 2x(1-x)\,r_\odot\,D_S\cos\ell\cos b}\,,
\end{equation}
where $r_\odot = 8.5\,$kpc and the distance to the source $D_s \simeq 50\,$kpc for LMC ($D_s \simeq 60\,$kpc for SMC). The longitude and latitude of the source in galactic coordinates are respectively $\ell = 280.46^\circ$ and $b = -32.89^\circ$ for LMC ($\ell = 302.81^\circ$ and $b = -44.33^\circ$ for SMC). In eq.~\eqref{eq:lensingevents}, several parameters are tailored to the EROS-2 survey. The instrument efficiency is taken from ref.~\cite{EROS-2:2006ryy}, the exposure time is $E \simeq 3.76\times 10^7$\,years ($E \simeq 5.9\times 10^6$\,years), the time range to cross the Einstein ring radius $t_{\rm E} \in [1,500]$, and the number of events is $N_{\rm events} = 3.9$. Results for the allowed fraction of galactic dense object are shown in figure~\ref{fig:lensingbound} for axion stars, which are generally too small to exhibit an extended distribution and thus appear in the survey as point-like objects.

\section{Conclusions}
\label{sec:conclusions}

In this paper, we have discussed the formation mechanism of a dark matter clump or dark ``minihalo'' around a primordial black hole (PBH) in the early Universe between the matter-radiation equality epoch at redshift $z_{\rm eq}$ and the onset of structure formation at $z_{\rm gal} \approx 30$. We have underlined the properties of such a dark minihalo, including its density profile, radius and mass $M_{\rm halo}$, according to previous studies. In the scenario where the dark matter is in the form of the QCD axion of mass $m_a$, we have considered the nucleation of axion stars inside an axion minicluster and a dark minihalo. We have extended the results obtained in previous literature to include the target of the axion star mass range that it available in the model, comparing it with current bounds from microlensing and theoretical findings for the axion mass. The results are shown in figure~\ref{fig:MASdensityplot} for the axion star mass in terms of the parameter space ($M_{\rm halo}, m_{\rm a}$).

By solving the Schr{\"o}dinger-Poisson equation in the dark minihalo background, we have obtained the profile and the mass-radius relations for a soliton forming inside such an environment, here referred to as a ``halo'' axion star. We remarked that the properties of this object depend on the central PBH mass as depicted in figure~\ref{fig:profileplot}. In particular, its non-trivial mass-radius relation $R_{\rm AS} = R_{\rm AS}(M_{\rm AS})$ differs from the power-law dependence $R_{\rm AS} \propto 1/M_{\rm AS}$ which has been historically observed for dilute axion stars in the vacuum. This has been explicitly shown in figure~\ref{fig:MRrelation} for axion miniclusters of different masses. The existence of a maximum stellar radius impacts on the mass distribution of these objects and it could be the subject for future numerical investigations, such as a systematic study of the condensation of these objects in halos of different masses.

To corroborate the study on possible indirect imprints, we have assessed the effects of axion stars onto gravitational microlensing. We have improved over previous studies on the QCD axion by deriving the lensing power of an axion star from the numerical resolution of the Schr\"odinger-Poisson system, as detailed in appendix~\ref{sec:microlensing}. We have tested our results against the analysis by the EROS-2 survey of the microlensing events observing both LMC and SMC, leading to the bounds on the fraction of axion stars shown in figure~\ref{fig:lensingbound}. While the results from the microlensing bounds match the existing limits from point-like lenses, we have shown that the numerical resolution of the equation supports the ansatz used in previous literature to model the inner profile of the axion star. Future work in this direction will focus on the inclusion of the halo mass distribution into the lensing bounds. The potential appearance of gravitational waves from binary mergers of halo axion stars in future setups has also been discussed. While the compactness of these objects might not be sufficiently large for the gravitational wave strain from galactic binaries to be currently detectable, some configurations could appear in the ongoing and upcoming searches.

\section*{Acknowledgements}

We thank E.\ Schiappacasse for reviewing the draft and for insightful comments. The authors acknowledge support by the National Natural Science Foundation of China (NSFC) through the grant No.\ 12350610240 ``Astrophysical Axion Laboratories''. L.V.\ thanks for the hospitality received by the Istituto Nazionale di Fisica Nucleare (INFN) section of Napoli (Italy), the INFN section of Ferrara (Italy), the INFN Frascati National Laboratories near Roma (Italy), the Galileo Galilei Institute for Theoretical Physics in Firenze (Italy), and the University of Texas at Austin (USA) throughout the completion of this work. This publication is based upon work from the COST Actions ``COSMIC WISPers'' (CA21106) and ``Addressing observational tensions in cosmology with systematics and fundamental physics (CosmoVerse)'' (CA21136), both supported by COST (European Cooperation in Science and Technology).

\appendix

\section{The inner structure of an axion star}
\label{sec:structureAS}

The dynamics of the axion field $\Phi$ under the influence of gravity is described by the action
\begin{equation}
	\label{eq:lagrangian}
	S = \int {\rm d}^4x \,\sqrt{-g}\left(\frac{1}{2} \left(\partial^\mu \Phi \right) \left(\partial_\mu \Phi\right) - V(\Phi/f)\right),
\end{equation}
where the metric $g^{\mu\nu}$ is determined by the Einstein equation for the energy momentum tensor of the axion field $T^{\mu\nu}(\Phi)$, and where $ V(\Phi/f)$ is the axion potential. We decompose the axion field as~\cite{Guth:2014hsa}
\begin{equation}
	\Phi = \frac{1}{\sqrt{2m}}\,\left[\psi({\bf r}, t)e^{-im t} + \psi^*({\bf r}, t)e^{im t}\right]\,,
\end{equation}
where $\psi$ is a wave function that describes the collective motion of a condensate formed with $N$ axions, normalized so that
\begin{equation}
	\label{eq:norm}
	\int {\rm d}^3 {\bf r}\,|\psi|^2 = N\,.
\end{equation}
Once the rapidly oscillating terms have been averaged out to zero, the kinetic term in eq.~\eqref{eq:lagrangian} reads
\begin{equation} \label{kinetic_term}
	\frac{1}{2} \left(\partial^\mu \Phi \right) \left(\partial_\mu \Phi\right)  = \frac{1}{2m}\,\dot{\psi}^*\,\dot{\psi} + \frac{i}{2}\,\left(\psi^*\,\dot{\psi} - \dot{\psi}^*\,\psi\right) -\frac{1}{2m}\,\nabla\,\psi^*\,\nabla\psi + \frac{m}{2}\,\psi^*\,\psi\,,
\end{equation}
where the term $\dot{\psi}^*\,\dot{\psi}/(2m)$ can be safely dropped, being much smaller than other terms in the non-relativistic limit. The self-interaction potential for the QCD axion is
\begin{equation}
	V_{\rm self}(\Phi) = \frac{1}{2}m^2\Phi^2 - \frac{\lambda_\Phi}{4!}\Phi^4\,,
\end{equation}
where $\lambda_\Phi = (1-3c_z)(m/f)^2$ and $c_z = z/(1+z)^2 \approx 0.22$. In the non-relativistic limit this reads
\begin{equation}
	V_{\rm self}(\Phi) = \frac{m}{2} \left| \psi \right|^2 -\frac{\lambda_\Phi}{16 m^2} \left| \psi \right|^4\,.
\end{equation}
The Lagrangian reduces to
\begin{equation}
	\label{eq:lagrangian0}
	\mathcal{L} = \frac{i}{2}\,\left(\psi^*\,\dot{\psi} - \dot{\psi}^*\,\psi\right) -\frac{1}{2m}\,\nabla\,\psi^*\,\nabla\psi + \frac{\lambda_\Phi}{16 m^2} \left| \psi \right|^4 - \frac{1}{2}m\phi|\psi|^2 \,,
\end{equation}
where the gravitational potential $\phi$ satisfies the Poisson equation
\begin{equation}
	\label{eq:poisson2}
	\nabla^2\,\phi = 4\pi G \left(\rho+\rho_{\rm PBH}\right)\,.
\end{equation}
Here, we have included the action of a PBH placed at the center of the distribution, with density $\rho_{\rm PBH} = M_{\rm PBH}\delta^{(3)}({\bf r})$. The Schr{\"o}dinger equation resulting from the Lagrangian above reads~\cite{Guzman:2004wj,Davies:2019wgi}
\begin{equation}
	\label{eq:schroedinger}
	i\frac{\partial \psi}{\partial t} = -\frac{1}{2m}\nabla^2\psi + m\phi\psi - \frac{\lambda_\Phi}{8m^2} |\psi|^2\psi\,,
\end{equation}
describing the motion of the axion under the influence of the self-interaction potential and gravity. The set of eqs.~\eqref{eq:schroedinger}-\eqref{eq:poisson} forms the Schr{\"o}dinger-Poisson system and it is analogous to the non-relativistic limit of the Gross-Pitaevskii equation when self-interactions are included~\cite{Chavanis:2011zi}. Setting
\begin{equation}
	\psi = e^{-i \epsilon m t}\,\sqrt{\rho(r)/m}\,,
\end{equation}
where $\epsilon$ is the binding energy per unit axion mass $m$ gives
\begin{eqnarray}
	\label{eq:quantum1}
	\rho''+\frac{2}{r}\rho' - \frac{1}{2\rho}(\rho')^2 &=& 4m^2\left(\phi - \epsilon - \frac{\lambda_\Phi}{8m^4}\rho\right)\rho\,,\\
	\phi''+\frac{2}{r}\phi' &=& 4\pi G\left(\rho+\rho_{\rm PBH}\right)\,.\label{eq:poisson1}
\end{eqnarray}
Once a solution for the density profile is found, the mass of the axion star is computed as
\begin{equation}
    M_{\rm AS} = \int_0^{+\infty} {\rm d}r\,4\pi r^2\,\rho(r)\,.
\end{equation}

\begin{figure}[htbp]
    \centering
    \includegraphics[width=0.7\linewidth]{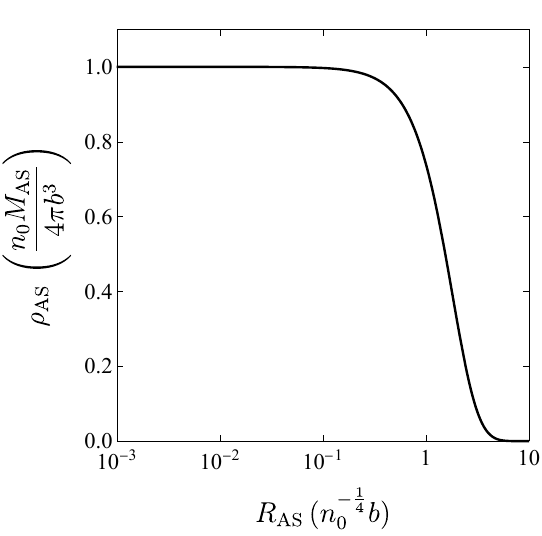}
    \caption{The density profile for a dilute axion star obtained from solving eq.~\eqref{eq:SchroedingerPoisson}, in the rescaled units reported in the axes.}
    \label{fig:profile}
\end{figure}

A radial equation for the density profile of the axion star is obtained by combining eqs.~\eqref{eq:quantum1} and~\eqref{eq:poisson1} as~\cite{Chavanis:2011zm}
\begin{eqnarray}
    \label{eq:schroedingerpoisson}
    &&32\pi G m_a^2 r \rho^4(\rho+\rho_{\rm PBH}) + 6 r \rho'^4-2\rho^3 \left(r \rho^{(4)} + 4 \rho^{(3)} \right)  - 2 \rho  \rho'^2\left(7 r \rho'' + 6 \rho' \right)\nonumber \\
    && + 2\rho ^2\left(2 r \rho''^2 +  \rho'\left(3 r \rho^{(3)} + 
       10 \rho'' \right) \right) +\frac{\lambda_\Phi}{m_a^2} \rho ^4\left(r \rho'' \ + 2 \rho' \right)=0\,,
\end{eqnarray}
where a prime is a derivation with respect to $r$. Here, the effect of the dark minihalo $\rho_H$ has also been included. We first neglect the contribution from $\rho_{\rm PBH}$ and we consider the solution for a dilute axion star in the vacuum. We also neglect the contribution of the self-interaction by setting $\lambda_\Phi = 0$. The boundary conditions for the differential equation are found by expressing $\rho$ as a Taylor series around $r=0$,
\begin{equation}
    \rho(r)=\sum_{i=0}^\infty a_i\,r^i\,,
\end{equation}
which, once plugged into eq.~\eqref{eq:schroedingerpoisson}, gives
\begin{equation}
     -6m^2 \rho'^3(0)+20 m^2 \rho(0)\rho'(0) \rho''(0)-24  m^2 \rho^2(0) \rho^{(3)}(0)=0\,.
\end{equation}
We look for a solution that possesses an inner core $\rho(0) \neq 0$. This leads to the requirements $\rho'(0) = \rho^{(3)} = 0$ while the values for $\rho(0)$ and $\rho'''(0)$ are yet to be determined. For this, the expression above is rescaled by introducing the length scale
\begin{equation}
    \label{eq:parameter_b}
 	b = (2GM_{\rm AS}\,m_a^2)^{-1}\,,
 \end{equation}
which corresponds to the size of the star under the gravitational pull. Setting $x = r/b$ and $R = 4\pi b^3\rho/M_{\rm AS}$ gives a differential equation for the rescaled profile $R$ with the initial condition $R = R_0$ at $x = 0$. Setting $X = R_0^{1/4}x$, $R = R_0 F$ allows us to recast eq.~\eqref{eq:schroedingerpoisson} in the form
\begin{equation}
    \label{eq:SchroedingerPoisson}
	F^{(4)} +\frac{4}{X}F^{(3)} - \frac{10F'' F'}{XF} + \frac{6(F')^3}{XF^2} - \frac{3F^{(3)} F'}{F} - \frac{2(F'')^2}{F} + \frac{7(F')^2F''}{F^2} - \frac{3(F')^4}{F^3} = 2F^2\,,
\end{equation}
with the initial conditions at $X = 0$ defined by $F=1$,  $F' = F^{(3)} = 0$, and $F''=F_2$, where the constant $F_2 = -0.61238693716$ is obtained via a shooting method. The normalization condition in eq.~\eqref{eq:norm} is recast as
\begin{equation}
	\label{eq:norm1}
	R_0 = \left(\int {\rm d}X X^2 \, F\right)^{-4} \approx 0.00691157\,,
\end{equation}
while the dimensionless radius that encloses 90\% of the mass is found as $X_{90} \approx 3.94166$. This translates into the radius
\begin{equation}
    R_{\rm AS} \approx \frac{6.84}{G\,m_a^2\,M_{\rm AS}}\,,
\end{equation}
while the radius that encloses 99\% of the axion star mass coincides with the results in previous work~\cite{Ruffini:1969qy, Membrado:1989bqo, Chavanis:2011zm}. Solving the equation above yields the density profile of the axion star shown in figure~\ref{fig:profile}. The density and the radius of the star are rescaled according to the parameter introduced in eq.~\eqref{eq:parameter_b}. 

\section{Microlensing of axion stars}
\label{sec:microlensing}

We summarise the equations used to derive the microlensing constraints in section~\ref{sec:discussion} for the abundances of axion stars and dark minihalos in the model considered. Additional details can be found in refs.~\cite{Narayan:1996ba, Dabrowski:1998ac, Fairbairn:2017sil, Croon:2020wpr, Croon:2020ouk, Fujikura:2021omw, Cai:2022kbp}. The detectability of a microlensing event from a sub-solar lens leads to a modification of the flux ${\cal F}_0$ in the absence of lensing by the quantity
\begin{equation}
    \mu \equiv {\cal F}/{\cal F}_0\,,
\end{equation}
where ${\cal F}$ is the observed flux.

The true source position angle $\beta$ with respect to the axis between the lens center and the source is determined by the path of light rays after a deflection from a massive lens of mass $M$ through the lensing equation
\begin{equation}
    \label{eq:lens}
    \beta=\theta-\frac{\theta_{\rm E}^2}{\theta}\frac{M(\theta)}{M}~,
\end{equation}
where $\theta$ is the angle of the observed lensed image of the source and the lens mass projected onto the lens plane is
\begin{equation}
    \label{eq:massproject}
    M_0(\theta) \equiv \int_0^r \,{\rm d}b\,2\pi b\,\int_{-\infty}^\infty {\rm d}z\,\rho\!\left(\sqrt{b^2+z^2}\right)\,.
\end{equation}
The Einstein angle appearing in eq.~\eqref{eq:lens} is defined as the solution of the lens equation when $\beta (\theta_{\rm E})= 0$ for a pointlike lens, as~\cite{Einstein:1936llh}
\begin{equation}
    \label{eq:rE} 
    \theta_{\rm E} \equiv\sqrt{4GM\frac{D_{\rm LS}}{D_\text{\tiny L}D_{\rm S}}} =\sqrt{\frac{4GM}{D_{\rm S}}\frac{1-x}{x}}\,,
\end{equation}
and the corresponding Einstein radius on the lens plane reads $R_{\rm E} \equiv D_{\rm L} \theta_{\rm E}$. Setting the angular distance from the lens center to the source center $u = \beta /\theta_{\rm E}$, the lensing eq.~\eqref{eq:lens} at the edge of the source is
\begin{equation}
    \label{eq:lens1}
    u = t - \frac{m(t)}{t}\,,
\end{equation}
where the rescaled projected mass for a spherically symmetric density profile $\rho(r)$ is
\begin{equation}
    \label{eq:massproject1}
    m(t) = \frac{\int_0^{t} {\rm d}\sigma \sigma \int_0^\infty {\rm d}\lambda\, \rho(R_{\rm E}\sqrt{\sigma^2+\lambda^2})}{\int_0^\infty {\rm d}\gamma \gamma^2 \rho(R_{\rm E}\gamma)}\,.
\end{equation}
Eq.~\eqref{eq:lens1} is solved to obtain the multiple positions of an image at the value $t_i(u) \equiv \theta_i/\theta_{\rm E}$. The magnification produced by an individual image $i$ is expressed by the ratio of the image area to the source area as~\cite{1994ApJ...430..505W}
\begin{equation}
    \label{eq:magfinitesource}
    \mu_i = \left|\frac{\theta}{\beta}\frac{\partial\theta}{\partial\beta}\right|\,,
\end{equation}
so that the total magnification $\mu_{\rm tot} = \sum_i\mu_i$ is the sum of the individual contributions. Finite source size effects are relevant when computing the magnification from lenses whose size is smaller than the wavelength of the detected light and dominate the suppression of lensing signatures for the lens masses $M \approx 10^{-11}\,M_\odot$. Here, we safely neglect these wave optics effects as discussed in previous work. In the limit of a negligible source size and pointlike lens, the lens eq.~\eqref{eq:lens1} is solved analytically to give
\begin{equation}
    \label{eq:mupoint}
    \mu_{\rm tot} = \frac{2+u^2}{u\sqrt{u^2+4}}\,.
\end{equation}
This gives the threshold luminosity $\mu_{\rm th} \approx 1.34$ when $u=1$, which corresponds to the threshold value adopted by the lensing surveys to define a detectable microlensing event. In the opposite limit of a very large source, the lensing only affects a negligible fraction of the light rays sourced so that eq.~\eqref{eq:lens1} predicts a large suppression of the luminosity.

\begin{figure*}[t!]
    \centering
    \begin{subfigure}[t]{0.32\textwidth}
        \centering
        \includegraphics[width=0.95\linewidth]{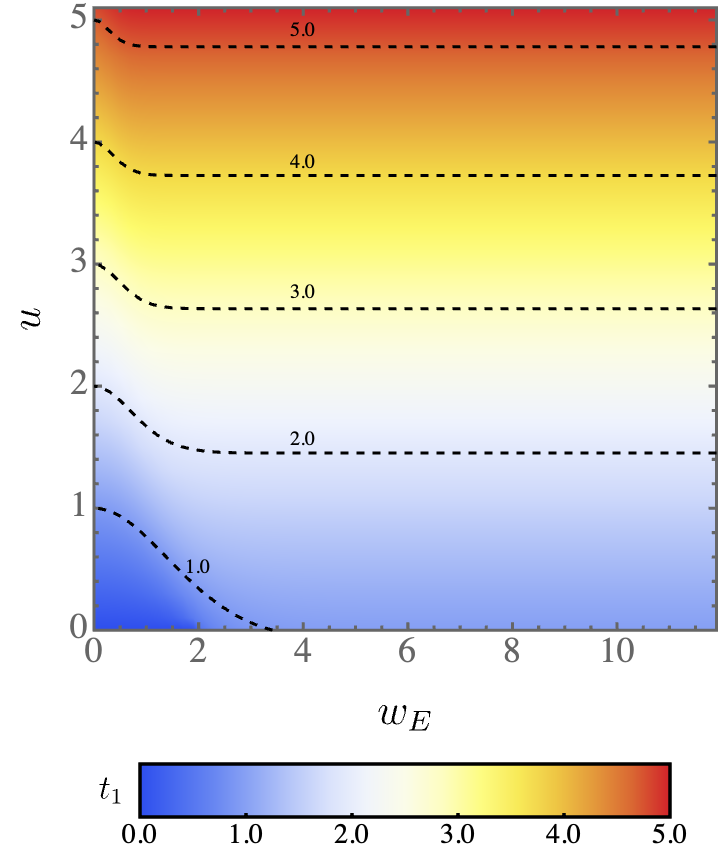}
    \end{subfigure}
    \begin{subfigure}[t]{0.32\textwidth}
        \centering
        \includegraphics[width=0.95\linewidth]{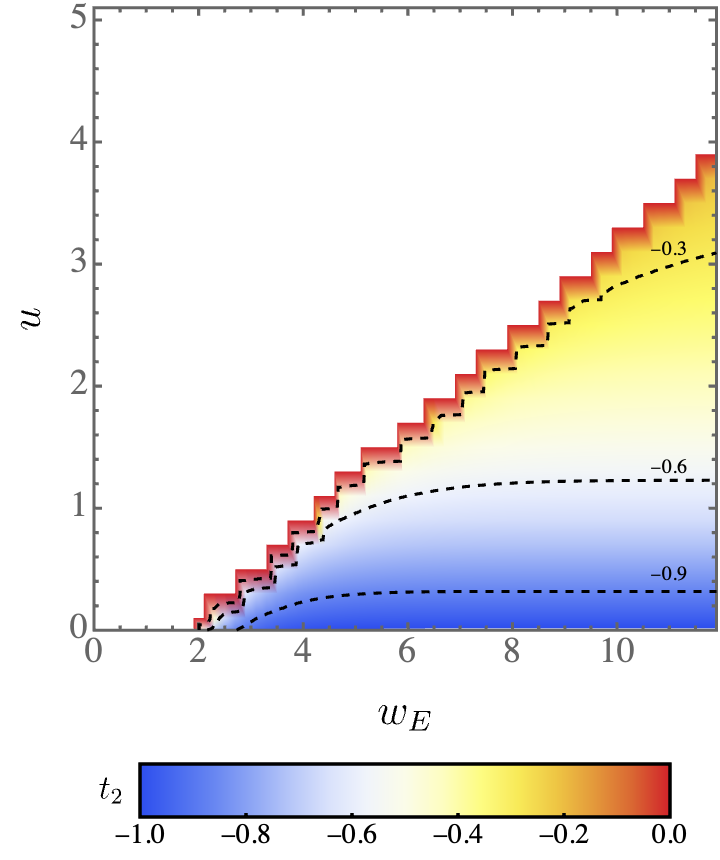}
    \end{subfigure}
    \begin{subfigure}[t]{0.32\textwidth}
        \centering
        \includegraphics[width=0.95\linewidth]{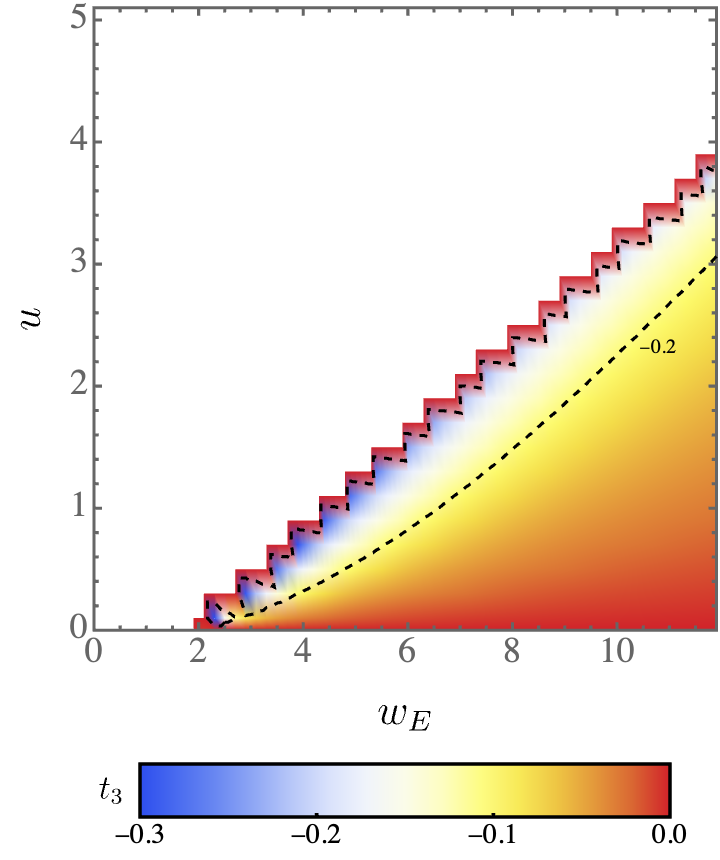}
    \end{subfigure}
    \caption{The solutions of the lens eq.~\eqref{eq:lens1}: $t_1$ (left), $t_2$ (middle), and $t_3$ (right), as a function of the impact parameter $u$ and the inverse extent of the lens $w_{\rm E}$.}
    \label{fig:tprofiles}
\end{figure*}

We derive the value for the threshold impact parameter $u_{1.34}$ that leads to the magnification $\mu_{\rm tot}> \mu_{\rm th}$ for two different systems, namely the axion star and the dark minihalo. For the case of the axion star, we employ the density profile that results from the numerical solution of Ees.~\eqref{eq:quantum}-\eqref{eq:poisson}, finding that there is a good overlap with the previous results where an analytical form is adopted~\cite{Fujikura:2021omw, Sugiyama:2021xqg, Schiappacasse:2021zlr}. This is shown in figure~\ref{fig:tprofiles} for the values of the three roots of eq.~\eqref{eq:lens1}, namely $t_1$ (left panel), $t_2$ (middle panel), and $t_3$ (right panel).

\begin{figure}[t!]
    \centering
    \includegraphics[width=0.7\linewidth]{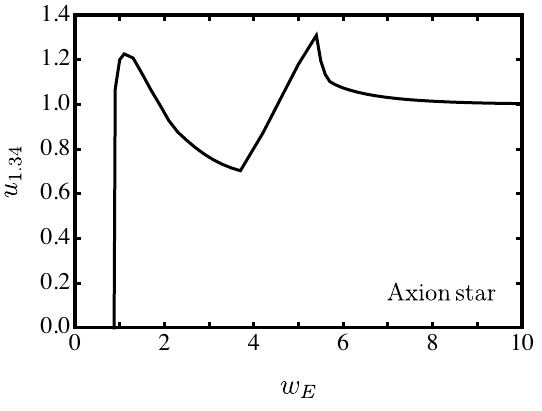}
    \caption{The threshold impact parameter $u_{1.34}$ as a function of $w_{\rm E} \equiv R_{\rm E}/R_{\rm AS}$ for an axion star.}
    \label{fig:u134}
\end{figure}

The numerical result is reported in figure~\ref{fig:u134} as a function of the ratio between the Einstein ring radius and the size of the lens, $w_{\rm E} \equiv R_{\rm E}/R_{\rm AS}$. The result coincides with the findings obtained through an analytical approximation in ref.~\cite{Fujikura:2021omw}, so we confirm the validity of such an approximation for the axion star profile.

\bibliographystyle{JHEP}
\bibliography{references.bib}

\end{document}